\documentclass[fleqn,usenatbib]{mnras}

\usepackage{newtxtext,newtxmath}

\usepackage[T1]{fontenc}
\usepackage{ae,aecompl}

\usepackage{graphicx}	
\usepackage{amsmath}	
\usepackage{amssymb}	

\title[Testing extreme-axion wave dark matter using the BOSS Lyman-Alpha forest data]{Testing extreme-axion wave-like dark matter using the BOSS Lyman-Alpha forest data}

\author[Ka-Hou Leong, Hsi-Yu Schive, Ui-Han Zhang, Tzihong Chiueh]{
Ka-Hou Leong$^{1}$, Hsi-Yu Schive$^{1,2}$, Ui-Han Zhang$^1$, Tzihong Chiueh$^{1,2,3}$\thanks{E-mail: chiuehth@phys.ntu.edu.tw}
\\
$^{1}$Physics Department, National Taiwan University, Taiwan\\
$^{2}$ Institute of Astrophysics, National Taiwan University, Taiwan\\
$^{3}$ Center for Theoretical Physics, National Taiwan University, Taiwan\\
}

\date{Accepted XXX. Received YYY; in original form ZZZ}

\pubyear{2015}

\begin{document}
\label{firstpage}
\pagerange{\pageref{firstpage}--\pageref{lastpage}}
\maketitle

\begin{abstract}
Using cosmological particle hydrodynamical simulations and uniform ultraviolet backgrounds, we compare Lyman-$\alpha$ forest flux spectra predicted by the conventional cold dark matter (CDM) model, the free-particle wave dark matter (FP$\psi$DM) model and extreme-axion wave dark matter (EA$\psi$DM) models of different initial axion field angles against the BOSS Lyman-$\alpha$ forest absorption spectra with a fixed boson mass $m_b\sim 10^{-22}$eV.  We recover results reported previously \citep{2017PhRvL.119c1302I,2017MNRAS.471.4606A} that the CDM model agrees better with the BOSS data than the FP$\psi$DM model by a large margin, and we find the difference of total $\chi^2$'s is $120$ for $420$ data bins.  These previous results demand a larger boson mass by a factor $>10$ to be consistent with the date and are in tension with the favoured value determined from local satellite galaxies.  We however find that such tension is removed as some EA$\psi$DM models predict Lyman-$\alpha$ flux spectra agreeing better with the BOSS data than the CDM model, and the difference of total $\chi^2$'s can be as large as $24$ for the same bin number.  This finding arises with no surprise since EA$\psi$DM models have unique spectral shapes with spectral bumps in excess of the CDM power near the small-scale cutoff typical of $\psi$DM linear matter power spectra as well as more extended cutoffs than FP$\psi$DM \citep{2017PhRvD..96b3507Z,2017PhRvD..96f3522Z}. 
%
%
\end{abstract}




\section{Introduction}

The remarkable success of Lambda cold dark matter model ($\Lambda$CDM) is a milestone in modern cosmology. Although the nature of dark energy and dark matter are still unknown, $\Lambda$CDM has been amply tested on many large-scale phenomena successfully, such as CMB fluctuations \citep{2016A&A...594A..13P}, baryon acoustic oscillations \citep{2007ApJ...664..675E}, accelerating expansion of the universe \citep{1998AJ....116.1009R,1999ApJ...517..565P}, etc.  Despite that, predictions of $\Lambda$CDM have however been in tension with observations on small scales ($\lesssim 10$ kpc). One well-known example is the missing satellites problem \citep{1999ApJ...522...82K}; a related problem is the too-big-to-fail problem \citep{2011MNRAS.415L..40B}; inside the dwarf spheroidal galaxies, there is a long-debated core-cusp problem \citep{1994Natur.370..629M}. This small-scale observational evidences cast doubt on the viability of the conventional particle cold dark matter (CDM) model.

Fuzzy dark matter \citep{2000PhRvL..85.1158H}, or wave dark matter \citep[$\psi$DM,][]{2014NatPh..10..496S} on the other hand provides a viable alternative solution to the small-scale problem \citep{2016PhR...643....1M}. The hypothesis made for $\psi$DM is that dark matter consists of extremely light bosons, $m_b \sim 10^{-22}$ eV. As $m_b$ is so small, quantum pressure arising from the uncertainty principle becomes manifestly effective on scales smaller than $10$ kpc and impacts on the cosmic structure, where small sub-halos are suppressed and cores of medium-size sub-halos smoothed. 
%
%
%
%

The wave dark matter model also has strong predictive power. It predicts that the first galaxy should form around $z=12$, the galaxy number count should be abruptly diminished beyond $z>9$ and the cosmic reionisation occurs late \citep{2016ApJ...818...89S}.  It also asserts every galactic halo of any mass should host one and only one stable high-density dark matter core (dubbed the soliton) and the mass of the soliton is strongly correlated with the mass of the halo \citep{2014PhRvL.113z1302S,2018arXiv180409647V}. The halo is composed of large-amplitude granules of roughly the same size fluctuating with roughly the same correlation time \citep{2018PhRvD..97j3523L}. The soliton can robustly survive even in the presence of more massive baryons often found in the inner halo of a galaxy \citep{2018MNRAS.478.2686C}, and can create detectable signatures in the core of Milky Way through pulsar timing \citep{2017PhRvL.119v1103D}.
%
%

Despite the initial success,  wave dark matter has recently faced a serious challenge from the Lyman-Alpha (Ly$\alpha$) forest observations \citep{2017MNRAS.471.4606A, 2017PhRvL.119c1302I}.  Models with boson masses of $1$ to few $10^{-22}$ eV determined by the soliton cores of satellite galaxies \citep{2014NatPh..10..496S,2017MNRAS.468.1338C} cannot reproduce the observed Ly$\alpha$ flux power spectrum; at least one order of magnitude higher boson mass is required to be consistent with the observations.  Not unlike warm dark matter, this challenge renders the wave dark matter model as an inconsistent model requiring a lower particle mass to be consistent with local satellite galaxy observations but at least a $10$ times higher particle mass for high-redshift Ly$\alpha$ forest observations. 

Recent theoretical developments have however discovered a possible solution for the dilemma faced by wave dark matter \citep{2017PhRvD..96b3507Z,2017PhRvD..96f3522Z} in the context of axion-like particles \citep{2017PhRvD..95d3541H}.  Having a capability to change the linear matter power spectrum shape, to which the Ly$\alpha$ power flux spectrum is most sensitive, this class of wave dark matter models, motivated by the QCD axion mechanism and called the extreme axion wave dark matter model (EA$\psi$DM), can provide a unique degree of freedom imprinted in the linear matter power spectrum not available to cold, warm or interacting particle dark matter models. 

The axion model has a field potential $V = m_b^2f^2(1 - \cos\theta)$, where $f$ is the axion decay constant or the axion symmetric breaking scale, and $\theta$ is the axion angle.  The axion potential becomes a simple harmonic oscillator when $\theta \rightarrow 0$, and we call this limit the free-particle $\psi$DM (or FP$\psi$DM). (In the non-relativistic limit, $m_b^2f^2\langle2\theta^2\rangle$ is the conventional mass density, where $\langle...\rangle$ is a short-time average to filter out the Zitterbewegung of rapid harmonic oscillation.) Due to the Hubble friction experienced by the axion field, any finite-amplitude initial $\theta$ will always approach this small-amplitude limit at a late time. If the initial $\theta$ is not only of finite amplitude but also very close to $\pi$, the unstable equilibrium, the angle $\theta$ will stay near the unstable equilibrium for a relatively long time, causing a delay in the nonlinear oscillation and producing parametric instabilities in the perturbation \citep{2017PhRvD..96f3522Z}. 

EA$\psi$DM can eventually become FP$\psi$DM in the matter-dominant era described by the Schroedinger-Poisson equation (with $Re[\psi]$ and $Im[\psi]$ to be identified as cosine and sine components of the Zitterbewegun oscillation, respectively); but EA$\psi$DM can have a different linear matter power spectrum from that of FP$\psi$DM, as the final spectrum depends on the spectral formation history mostly prior to the radiation-matter equality.  This difference in the linear matter spectrum can significantly alter the predicted Ly$\alpha$ flux power spectrum. In this paper, we focus on testing the EA$\psi$DM model against the Ly$\alpha$ flux power spectra obtained from the BOSS survey \citep{2013A&A...559A..85P}, and conclude that in some narrow range of initial angle of $\theta$, the trouble faced by the wave dark matter can be alleviated.

This paper is organised as follows. In Sec. \ref{sec:methodology}, we describe our simulations. We analyse the matter power spectra and compare flux power spectra with the BOSS Ly$\alpha$ data in Sec. \ref{sec:results}. In Sec. \ref{sec:discussion}, we discuss our findings in the comparisons with the BOSS data. Finally, we present our conclusion in Sec. \ref{sec:conclusion}. Appendix \ref{sec:Splicing} describes the splicing method. Appendix \ref{sec:convergence} shows the convergence test of flux power spectrum. We also present a comparison of representative Ly$\alpha$ spectra of different models in Appendix \ref{sec:spectra}. The suffixes "c" in length scales refer to units in the co-moving coordinate. 



\section{METHODOLOGY}
\label{sec:methodology}

\subsection{Power Spectra}
\label{sec:Power_Spectra} 

\begin{figure}
	\includegraphics[width=\columnwidth]{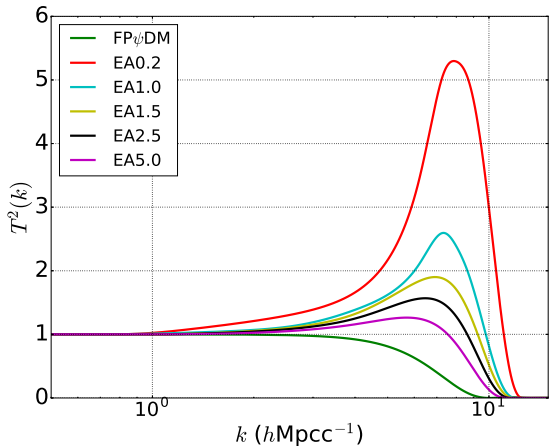}
    \caption{ $\psi$DM-to-CDM transfer functions for FP$\psi$DM, EA0.2, EA1, EA 1.5, EA 2.5 and EA 5.0 at z=49 with $m_{22} = 1.1$. Note that all $\psi$DM models have strong spectral suppression at high $k$, but only EA$\psi$DM models have broad spectral bumps.}
    \label{fig:transfer_function}
\end{figure}

One of the common features of wave dark matter power spectra is the suppression of small-scale structures below the Compton length prior to the matter-radiation equality \citep{2000PhRvL..85.1158H,2017PhRvD..96b3507Z} and the existence of a Jeans length due to the uncertainty principle in the matter dominant era \citep{2009ApJ...697..850W}.  The EA$\psi$DM power spectra, however, have one more degree of freedom other than the boson mass $m_b$, the initial misaligned field angle $\theta$, which gives rise to a broad spectral bump immediately longward of the spectral suppression (see Fig. \ref{fig:transfer_function}).  The bump and suppression features are best shown in the $\psi$DM-to-CDM transfer function, $T^{2}_{\psi \rm DM}(k, z) = P_{\psi \rm DM}(k, z)/P_{\rm CDM}(k, z)$, where $P$ is the matter power spectrum. Although $T^{2}_{\psi \rm DM}(k, z)$ generally depends on redshifts, \cite{2017PhRvD..96f3522Z} showed that the dependence of the CDM-to-$\psi$DM transfer function on redshift $z$ is extremely weak in the wavenumber regime probed by Ly$\alpha$ observations ($k \lesssim 5.9$ $h$Mpcc$^{-1}$). Specifically, a recent study \citep{2017MNRAS.471.4606A} discussed that the quantum pressure influence on the $\psi$DM dynamics at wavenumbers probed by Ly$\alpha$ forest observations could be neglected in the particle mass range $m_{22}(\equiv m_b/10^{-22}$ eV) $\gtrsim 1.1$.

To set the stage for this investigation, we show in Fig. \ref{fig:transfer_function} the transfer functions $T^{2}_{\psi \rm DM}$ of FP$\psi$DM, EA$\psi$DM with $\delta\theta(\equiv\pi-\theta)=0.2^{\circ}$, $1^{\circ}$, $1.5^{\circ}$, $2.5^{\circ}$ and $5^{\circ}$ at $z=49$. (Hereafter, we refer to EA$\psi$DM, for example, with $\delta=0.2^{\circ}$ as EA0.2.) From Fig. \ref{fig:transfer_function}, it is clear that the smaller the $\delta\theta$ the larger the cut-off wavenumbers $k_c$ (defined by T($k_c$) = 0.5). These EA$\psi$DM models have spectral bumps immediately longward of the cut-off, suggesting that the halo assembly history in EA$\psi$DM can significantly deviate from the traditional CDM and the FP$\psi$DM predictions at high redshifts \citep{2018MNRAS.473L..36S}. Therefore, the Ly$\alpha$ forest measured in the redshift range $z\sim 2-4$ can impose constraints on the particle mass and the initial field angle of EA$\psi$DM. 

In fact, the  Ly$\alpha$ flux power spectrum is affected by the cut-off $k_c$ of the linear matter power spectrum even more sensitively than the nonlinear matter power spectrum at $z\sim 2-4$. The former probes an intermediate gas density up to unity optical depth and is thus sensitive to the cosmic web, a structure in the relatively weakly nonlinear regime.  By contrast, the matter power spectrum is dominantly contributed by the highly nonlinear collapsed halos. In this regard, the quasar flux spectrum is less affected by nonlinear evolution and capable of capturing the linear matter power spectrum.

\subsection{Hydrodynamical simulations}
\label{sec:Hydrodynamical} 

Accurately simulating $\psi$DM dynamics needs to solve the Schr\"{o}dinger-Poisson equation \citep[e.g.][]{2014NatPh..10..496S,2014PhRvL.113z1302S}. However, such simulations in large ($\sim (100$ $h^{-1}$Mpcc $)^3$) simulation boxes with sufficiently high ($\sim 100$ $h^{-1}$pcc ) resolution are prohibitively computationally expensive even with the most advanced supercomputers. This is due primarily to that the time-step of computation scales unfavourably with the squared grid size in a zoom-in calculation in order to capture quantum pressure inside collapsed halos.  Hence, it is reasonable to inquire under what conditions CDM-hydro simulations with the initial matter power spectrum modified to the $\psi$DM spectrum can still capture the Ly$\alpha$ forest in the $\psi$DM scenario.  It turns out that the relatively low-resolution dynamical range of the BOSS data can validate this approach. (See Sec \ref{sec:Matter_power_spectrum} for details.)   By contrast, high-resolution observations, such as the XQ-100 \citep{2017MNRAS.466.4332I} and HIRES/MIKE \citep{1994SPIE.2198..362V,2002SPIE.4485..453B}, measure scales down to $100$ $h^{-1}$kpcc, a scale that is well within the quantum suppression regime, and hence CDM-hydro simulations are not valid. Even for CDM such high-resolution simulations are technically very challenging. \citep[See][ which discussed high accuracy Ly$\alpha$ forest simulations of the CDM model]{2015MNRAS.446.3697L}. In an attempt to solve the quantum pressure problem for high-resolution simulations, a modified CDM-hydro code (AX-GADGET) has been developed, where a quantum force law has been implemented in every fluid element \citep{2018MNRAS.478.3935N}. However, it is not clear to what extent such a modified hydro scheme can capture physics inside halos within $100$ kpc scale.

Various previous works have adopted the CDM-hydro approach with a modified matter spectrum to approximate wave dark matter dynamics.  \cite{2016ApJ...818...89S} pointed out that the growth rate ratio between $\psi$DM and CDM, defined in equation (4) of \cite{2016ApJ...818...89S} for $m_{22} = 1.6$, is almost unity for $k \leqslant 11$ $h$Mpcc$^{-1}$. Using a particle-mesh scheme, \cite{2016PhRvD..94l3523V} showed the quantum pressure effect is not apparent for FP$\psi$DM matter power spectrum until $k \gtrsim 150$ $h$Mpcc$^{-1}$ for $m_{22} = 2.5$. \cite{2017MNRAS.471.4606A} also suggested that quantum pressure is likely negligible for Ly$\alpha$ forest simulations with $m_{22}\gtrsim 1$.  The conclusions of these works are consistent with the approach we adopt.  For these very reasons, we shall be contented with the low-resolution BOSS data with CDM-hydro simulation predictions.

In this work, we perform $N$-body hydro simulations to represent $\psi$DM dynamics.  The simulation uses the mesh-free hydrodynamic simulation code, \textbf{GIZMO} \citep{2015MNRAS.450...53H}, which adopts the Lagrangian meshless finite mass (MFM) algorithm. Comparing with the traditional SPH method, MFM has many advantages. The most noticeable of all is that MFM does not require artificial viscosity, an infamous feature adopted by the SPH method to make computation stable. We use the \textbf{MUSIC} code \citep{2011MNRAS.415.2101H} to generate the initial conditions at $z = 49$ with the second-order Lagrangian perturbation theory method. The CDM power spectrum is generated by the \textbf{CAMB} package \citep{Lewis:2002ah}, the FP$\psi$DM transfer function is generated by the \textbf{AxionCAMB} \citep{2017PhRvD..95l3511H}, and for EA$\psi$DM transfer functions we follow the work of \cite{2017PhRvD..96f3522Z}. The chemistry and cooling library \textbf{GRACKLE} \footnote{https://grackle.readthedocs.io/}\citep{2017MNRAS.466.2217S} is used to solve the radiative processes. Also, all the matter power spectra are computed by \textbf{GenPK} \citep{2017ascl.soft06006B}

To capture the signal of the 1D flux power spectrum from small to large scales, we perform 3 simulations with $(L, N) =  (25$ $h^{-1}$Mpcc, $2\times512^{3})$, $(25$ $h^{-1}$Mpcc, $2\times128^{3})$ and $(100 $ $h^{-1}$Mpcc$, 2\times512^{3})$, where $L^3$ is the co-moving volume of the simulation box and $N$ is the total number of all particles (dark matter and gas). We apply a simple star formation criterion --- gas particles satisfying temperature $T < 10^5 $K and overdensity $\Delta > 1000$ are transformed to collisionless stars \citep{2004MNRAS.354..684V,2010JCAP...06..015V}.  We do not consider metal cooling because the metal abundances in IGM is negligible. The gravitational softening length for three different species (dark matter, gas and stars) are the same, set to $1/25$ of the mean co-moving interparticle distance of dark matter particles. The kernel of MFM is cubic spline and the effective neighbour of the kernel is $32$.  In all simulations, the primordial helium mass fraction is $Y = 0.24$.  This set of parameters is the same as most previous works on the Ly$\alpha$ flux power spectrum \citep[e.g.][]{2014JCAP...07..005B,2015JCAP...02..045P,2017MNRAS.464..897B}.

In focusing on the difference produced by different initial matter power spectra of CDM, FP$\psi$DM and EA$\psi$DM, we perform all simulations with identical Gaussian random seeds, cosmological parameters and ultraviolet background (UVB). The adopted cosmological parameters are the best fit result for the CDM model suggested by \cite{2015JCAP...02..045P}, where $\Omega_{m} = 0.292$,  $\Omega_{b} = 0.050$, $\sigma_{8} = 0.858$, $h = 0.668$, $n_{s} = 0.929$, which are fixed for all dark matter models. We also fix the $\psi$DM  mass parameter $m_{22} = 1.1$. We use the homogeneous intergalactic UV background of \cite{2012ApJ...746..125H}, assuming the intergalactic gas is highly ionised, in ionisation equilibrium and optically thin for photoionisation and photoheating. We also follow a modification introduced by \cite{2017MNRAS.464..897B}, the Sherwood simulation, to the HeII photoheating rate to ensure the IGM temperature in simulations is in agreement with the IGM temperature estimated from observations \citep[e.g.][]{2011MNRAS.410.1096B}. The modification adopts $\epsilon_{\rm He II} = 1.7\epsilon_{\rm He II}^{\rm HM12}$ from $z = 2.2$ to $z = 3.4$. This boosting factor can be regarded as an approximation method \citep{1998MNRAS.301..478T} to radiative transfer and non-equilibrium effects during HeII reionisation \citep{2015MNRAS.450.4081P}.  In low-density regions, the gas density and temperature are closely related, where the thermal behaviour of gas is dominated by adiabatic expansion cooling and photoionisation heating. The density-temperature relation for low density gases can roughly be expressed by a redshift-dependent power law, $T(z)=T_{o}(z)\Delta^{\gamma(z)-1}$, where $T_{o}$ is the mean temperature of low density IGM and $\Delta = \rho /  \langle\rho \rangle$ is the gas density over the background gas density. In this work, we find $T_{o}(z= 3.0) \simeq 12,500$ K and $\gamma(z=3.0) \simeq 1.54$ for all simulations, consistent with the simulation work of \cite{2017MNRAS.464..897B}.
 
\subsection{Mock Lyman-Alpha forest spectra}
\label{sec:Mockspectra} 

We extract various quantities, such as the velocity field, the internal energy and the neutral hydrogen density from \textbf{GIZMO} snapshots to calculate mock absorption Ly$\alpha$ spectra. To avoid periodical signals, we construct $10^{5}$ lines of sight (LOS) with random origins and directions at each redshift interval of $\Delta z = 0.2$ for all observed redshifts. The transmitted flux is defined as $F (v_{j})= e^{-\tau(v_{j})}$, where $\tau(v_{j})$ is the optical depth of Ly$\alpha$ absorption at the velocity coordinate $v_j$ (equivalent to the wavelength coordinate defined below). The optical depth $\tau$ along each LOS is computed through the Voigt-Hjerting function, expressed as  \citep[see][for more details]{2006MNRAS.369.2025T}:

\begin{equation}
   \tau(v_j) = \Delta x \sigma_{\alpha} c \sum_{i}^{N_{\rm LOS}} \frac{n_i}{\sqrt{\pi}v_{i}^{th}} \exp(-\frac{(v_{j}-(v_{i}^{H}+v_{i}^{pec}))^2}{({v_{i}^{th}})^2})
	\label{eq:delta_optical}
\end{equation}
where $\Delta x$ is the pixel length, the Ly$\alpha$ cross-section $\sigma_{\alpha} = 4.45 \times 10^{-18}$ cm$^2$, $c$ is the speed of light,  $j$ is the $j^{th}$ pixel on the LOS, $n_i$ the number density of neutral hydrogen at pixel $i$, $N_{\rm LOS}$ the total number of pixels on the LOS, the Doppler velocity $v_{j}= c(\frac{\lambda_{\rm Ly \alpha} - \lambda_{j}}{\lambda_{j}})$, $\lambda_{\rm Ly\alpha}=1216$ \AA, the thermal velocity $v_{i}^{th}=\sqrt{\frac{k_{b}T_{i}}{m_p}}$, $k_{b}$ is the Boltzmann constant and the Hubble velocity $v_{i}^{H} = i\Delta x H$. The 1D flux power spectrum $P_{F}^{1D} = \langle \left| FT[\delta_{LOS}] \right|^{2} \rangle$, the ensemble average of the squared Fourier amplitude of the transmitted flux fraction over different LOS's, where $\delta_{LOS} = F/\langle F \rangle - 1$. To ensure the simulated $P_{F}^{1D}$ is numerically convergent, the pixel length of spectra is set to be $0.69 $ km s$^{-1}$ which is $100$ times smaller than pixel size in the BOSS data. 

\cite{2014JCAP...07..005B} posited that large-volume, high resolution hydrodynamical simulations are necessary for simulating $P_{\rm F}^{\rm 1D}$ in order to cover the dynamical range of the BOSS data, $k = 1\times10^{-3} - 2\times10^{-2} $ s km$^{-1}$. Such a high dynamical range hydro simulation is computationally expensive. To achieve a sufficient dynamic range within reasonable time, we follow the splicing method suggested in \cite{2003ApJ...585...34M} and \cite{2014JCAP...07..005B} for constructing the flux power spectrum by combining large-scale and small-scale spectra. The key assumption of the splicing method is the ratio of high mass resolution $P_{\rm F}^{\rm 1D}$ to low mass resolution $P_{\rm F}^{\rm 1D}$ depends only on mass resolution of simulation. In our setting, this assumption can be succinctly written as:
\begin{equation}
    \frac{P_{\rm F,100,2048}^{\rm 1D} (k)}{P_{\rm F,25,512}^{\rm 1D} (k)}= \frac{P_{\rm F,100,512}^{\rm 1D} (k)}{P_{\rm F,25,128}^{\rm 1D} (k)}
	\label{eq:Splicing_I}
\end{equation}
In equation~(\ref{eq:Splicing_I}), the mass resolutions in the numerator and in the denominator are the same on either side. The splicing method makes use of equation~(\ref{eq:Splicing_I}) to merge 1D flux power spectra of different resolutions and different box sizes into a large-volume, high-resolution $P_{\rm F,100,2048}^{\rm 1D} (k)$ flux spectrum. In Appendices \ref{sec:Splicing} and \ref{sec:convergence}, we present how well the splicing method is justified. 

\subsection{Fitting parameters}

\label{sec:Fitting} 
We consider two categories of fitting parameters in the comparison procedure after the 1-D flux power spectrum is obtained. The first category applies different global effective optical depth, $\tau_{\rm eff}(z)\equiv-\ln(\langle F\rangle(z))$, to the simulation flux power spectrum $P_{\rm F}^{\rm 1D, th} (k)$. The second category takes into account observational imperfections of the BOSS data and simulations.

\begin{figure}
	\includegraphics[width=\columnwidth]{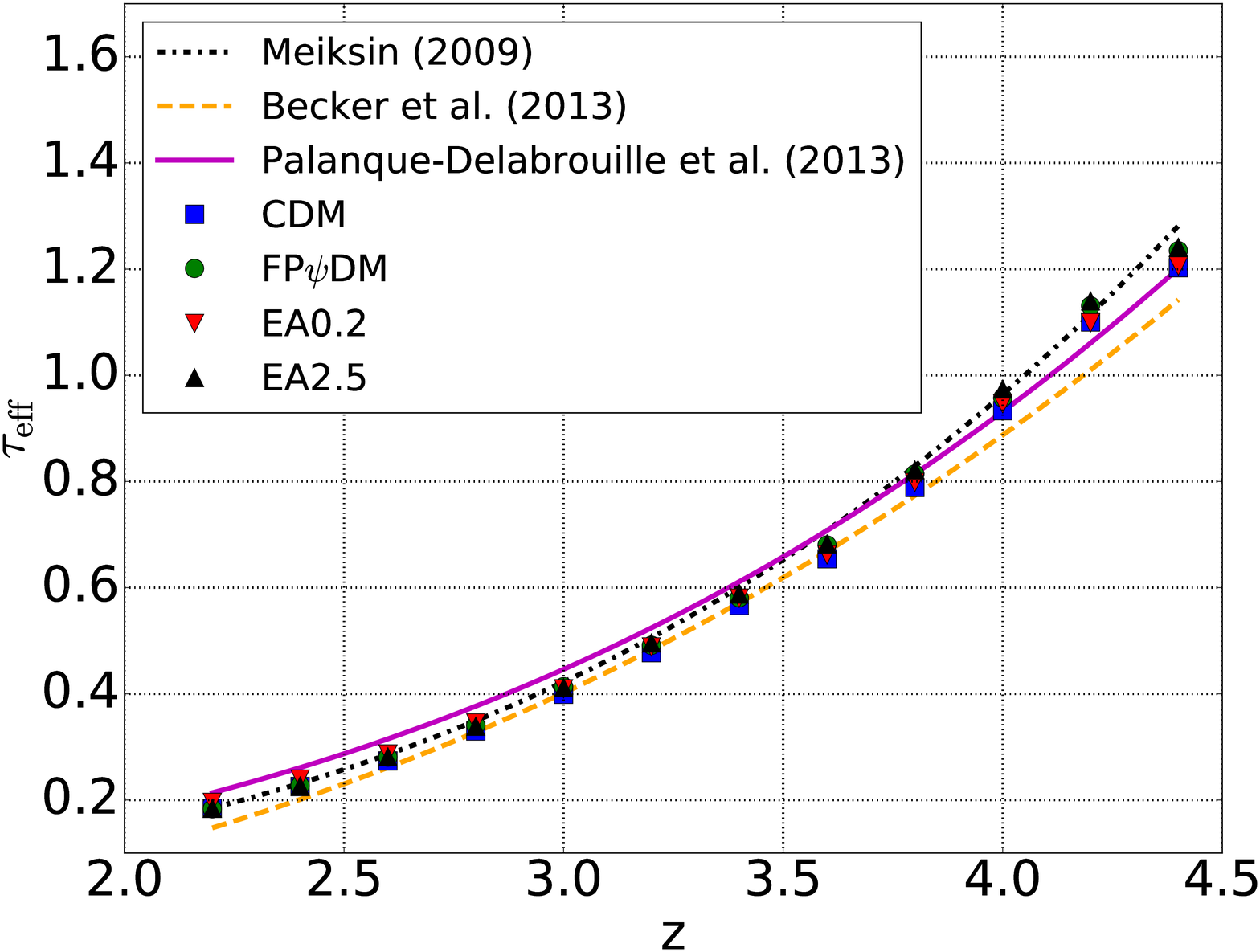}
    \caption{The best-fit $\tau_{\rm eff}$'s of CDM, FP$\psi$DM, EA0.2 and EA2.5 simulations at each redshift bin. The lines are three different empirical power law of $\tau_{\rm eff}$ \citep{2009RvMP...81.1405M, 2013MNRAS.430.2067B, 2013A&A...559A..85P}. Note that the best-fit $\tau_{\rm eff}$ of all dark models are similar to each other and consistent with the lines.}
    \label{fig:Optical_Depth}
\end{figure}

\begin{enumerate}
\item \textbf{Astrophysical parameter:} 

The effective optical depth $\tau_{\rm eff}$ is related to the photoionisation rate in IGM. In most simulation works \citep[e.g.][]{1998MNRAS.301..478T,2014JCAP...07..005B,2017MNRAS.464..897B,2017MNRAS.471.4606A}, $\tau_{\rm eff}$ is rescaled to follow a power law, $\tau_{\rm eff}(z) = A(1+z)^{B}$, where $A$ and $B$ are constants determined from the observational data.  In this work, we do not demand $\tau_{\rm eff}(z)$ to follow a particular empirical power law, for a reason that parameters $A$ and $B$ are different in different observations \citep[e.g.][]{2009RvMP...81.1405M,2013MNRAS.430.2067B,2013A&A...559A..85P}.  As a result, we choose to independently adjust $\tau_{\rm eff}(z)$ at each redshift for improving the goodness of fit between the predicted $P_{\rm F}^{\rm 1D, th} (k)$ and the BOSS data. In Fig. \ref{fig:Optical_Depth} we show the best-fit $\tau_{\rm eff}$, and they are found to lie between bounds of the empirical fittings of different observations. 

\item \textbf{Technical parameters:}
We next consider the second category of fitting parameters including three factors of technical origins.   These factors are contamination of SiIII and Ly$\alpha$ cross-correlation $C_{\rm Si}$, imperfection in noise estimate in the BOSS data $C_{\rm noise}$ and imperfection of the simulation resolution $C_{\rm reso}$. The impacts from the three factors to the predicted spectrum are expressed by following formula:

\begin{equation}
  P_{\rm F}^{\rm 1D, fit}  = P_{\rm F}^{\rm 1D, th} \times (C_{\rm Si}(k, z) \times C_{\rm reso}(k)) + C_{\rm noise}(k, z) 
	\label{eq:correlation}
\end{equation}
\begin{itemize}
\item \textbf{$C_{\rm Si}(k, z_i)$: }Due to the juxtaposition of the two lines, $\lambda_{Ly\alpha} = 1216$ {\AA} and $\lambda_{\rm SiIII} = 1206$ {\AA}, there is contamination from the SiIII absorption line to the Ly$\alpha$ absorption line in the observed $P_{\rm F}^{\rm 1D} (k)$ \citep{2006ApJS..163...80M,2013A&A...559A..85P}.  We adopt a multiplicative term, $C_{\rm Si}(k, z) = (1+a\cos{kv})^2+(a\sin{kv})^2 = 1 + a^2 + 2a\cos{kv}$, to the predicted flux power spectrum first introduced in \cite{2006ApJS..163...80M} to account for such contamination. Here $a = f/(1-\langle F \rangle(z))$ with f being a redshift-independent fitting parameter and $v = 2270$ km/s.
\item \textbf{$C_{\rm reso}(k)$: }We consider the imperfect resolution of simulation which possibly affects our comparison, and allow for a redshift-independent multiplicative correction factor, $C_{\rm reso}(k) = \exp{(-k^2 \cdot \alpha_{\rm reso})}$, as described in \cite{2015JCAP...02..045P}, where $\alpha_{\rm reso}$ is a random number obeying a zero-mean Gaussian distribution with a variance $\sigma = 5$ and its amplitude is a redshift-independent fitting parameter.
\item \textbf{$C_{\rm noise}(k, z)$: }The errors of measurement noise estimation need also to be accounted for. Following the method described in \cite{2015JCAP...02..045P}, we allow for $\pm 10 \%$ rms errors in the observation noise power spectra at each redshift and include an additive correction term in each redshift bin, $C_{\rm noise}(k, z) = P_{\rm noise}(k, z) \cdot \alpha_{\rm noise}(z)$, where $\alpha_{\rm noise}(z)$ is a random number obeying a zero-mean Gaussian distribution with a variance $\sigma = 0.1$.  Again its amplitude is a fitting parameter dependent on redshifts.  Here the noise power spectrum $P_{\rm noise}(k, z)$ has been given in the Boss data.  

In total we have $26$ fitting parameters for a sample of $420$ Boss data of all redshifts.  These fitting parameters are independently adjusted so as to reach a global best fit for each model. We want to stress that, except for $\tau_{\rm eff}(z)$, this work follows all remaining $14$ fitting parameters formulated in \cite{2006ApJS..163...80M} and \cite{2015JCAP...02..045P}.
\end{itemize}
\end{enumerate}

\section{RESULTS}
\label{sec:results}
\subsection{Matter power spectrum}
\label{sec:Matter_power_spectrum} 
\begin{figure*}
	\includegraphics[width=\textwidth]{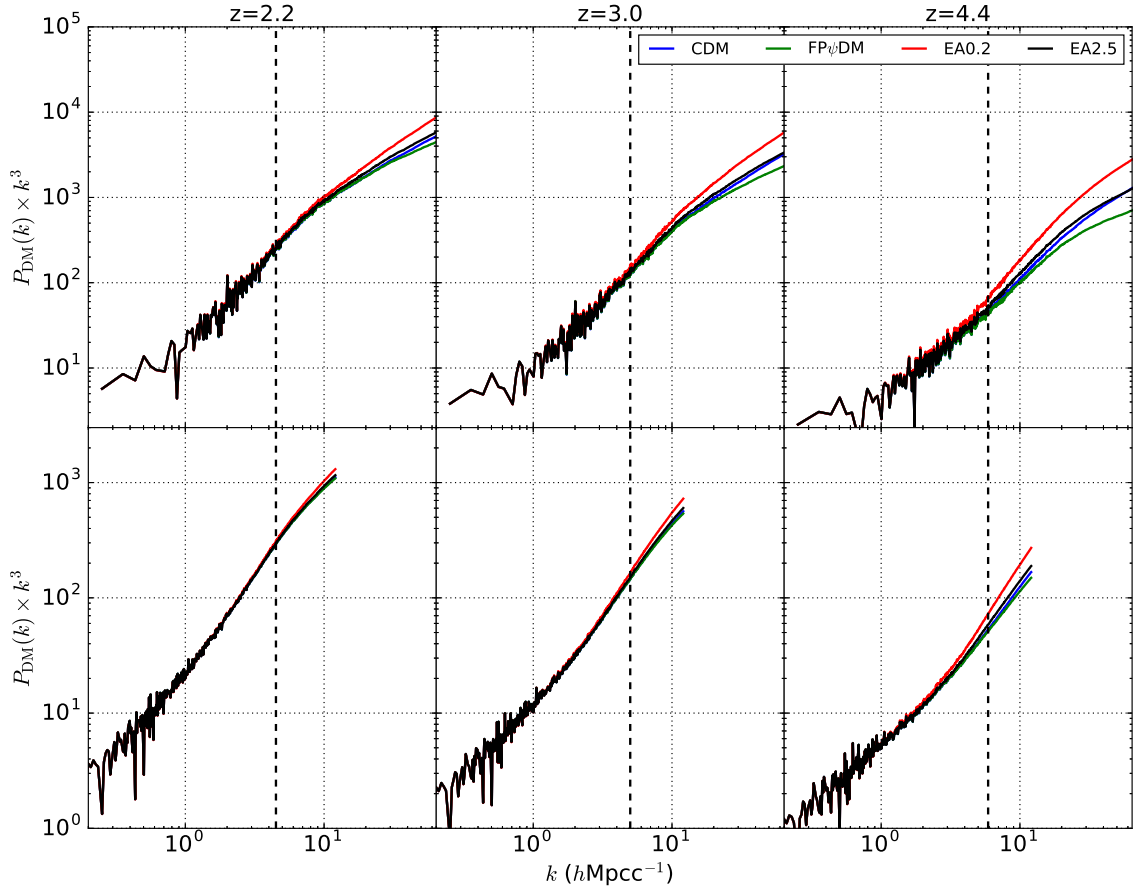}
    \caption{Matter power spectra of different dark matter models at z=2.2, 3.0 and 4.4. The difference in high-$k$ among 4 simulations reveals the effects of simulation resolution and nonlinearity on the shape of transfer functions. The dashed lines mark the Nyquist $k$ of the BOSS data $k_{Ny, BOSS}$. Upper panels: results from ($25$ $h^{-1}$Mpcc, $2 \times 512^3$) simulations. Lower panels: results from ($100$ $h^{-1}$Mpcc, $2 \times 512^3$) simulations.}
    \label{fig:DM_PS_I}
\end{figure*}

We first present the matter power spectrum of dark matter in co-moving coordinate at z=2.2, 3.0 and 4.4 with identical cosmological parameters and $m_{22} = 1.1$. Fig. \ref{fig:DM_PS_I} shows nonlinear matter power spectra of 6 different dark matter models from $(25$ $h^{-1}$Mpcc, $2\times512^{3})$ simulations and $(100$ $h^{-1}$Mpcc, $2\times512^{3})$ simulations. They are indistinguishable at low-$k$ and diverging at high-$k$.  The matter power spectra of all EA$\psi$DM are always larger than those of FP$\psi$DM at high-$k$, and the result shows that the small-scale suppression in the initial matter power spectra of EA$\psi$DM has been erased to various degrees, and they are replaced by nonlinear cascade spectra from the spectral bumps on the intermediate scale after evolution.   Despite the nonlinear effect, this comparison of different models provides an intuitive understanding of our model predictions of Ly$\alpha$ flux spectra to follow. 

In the absence of quantum pressure in simulations, the small-scale power of $\psi$DM simulations will certainly  be overestimated around $k_{\rm Ny,25} = 512\pi/25$ $h$Mpcc$^{-1}$ $\simeq 64$ $h$Mpcc$^{-1}$, the Nyquist frequency of $(25$ $h^{-1}$Mpcc, $2\times512^{3})$ simulations, as the quantum pressure becomes effective when $k $  is greater than the Jeans wavenumber $k_{J,\psi}$ of $\psi$DM.   Before the cosmological constant sets in to affect the Hubble expansion, $k_{J,\psi}$ is given by the following equation \citep{2009ApJ...697..850W}:

\begin{equation}
    k_{J,\psi} = (6a)^{1/4}(\frac{m_bH_0}{\hbar})^{1/2}(\Omega_{m})^{1/4},
    	\label{eq:Jean_length}
\end{equation}
where $a$ is the scale factor, $H_0$ is the Hubble constant, $\Omega_{m}$ is the current matter density and $\hbar$ is the reduced Planck constant.  We estimate $k_{J,\psi} \simeq 32 \sim 28$ $h$Mpcc$^{-1}$ in the range $z \simeq 2.2 - 4.4$ using equation~(\ref{eq:Jean_length}), hence $k_{\rm Ny,25} > k_{\rm J,\psi}$ and the high-$k$ spectrum is over-estimated, consistent with the results of \cite{2016PhRvD..94l3523V}. On the other hand, the maximal resolution of BOSS Ly$\alpha$ data is 69 km/s, corresponding to $k_{\rm Ny,BOSS} \simeq 4.5  - 5.9$ $h$Mpcc$^{-1}$ in the range $z \simeq 2.2 - 4.4$ (marked in Fig. \ref{fig:DM_PS_I}), and so $k_{\rm Ny,BOSS} \ll k_{J,\psi}$. Furthermore, limited by instrumental noise, the maximum usable $k$ of the BOSS data is $k_{\rm max, BOSS} \simeq 1.7 - 2.2$ $h$Mpcc$^{-1}$ in the range $z \simeq 2.2 - 4.4$, i.e., $k_{\rm max, BOSS}\sim 0.4 k_{\rm Ny, Boss}$, which is the highest wavenumber of our comparison. Thus, the dynamical range which BOSS Ly$\alpha$ measures cannot to be affected by the quantum pressure, and it is safe to assert that our simulations be reliable and the simulated $P_{\rm F}^{\rm 1D}$ accurately captures the wavenumber range of BOSS spectra. 

\subsection{Comparison simulated Lyman-Alpha flux power spectrum with BOSS data}

\begin{figure*}
	\includegraphics[width=\textwidth]{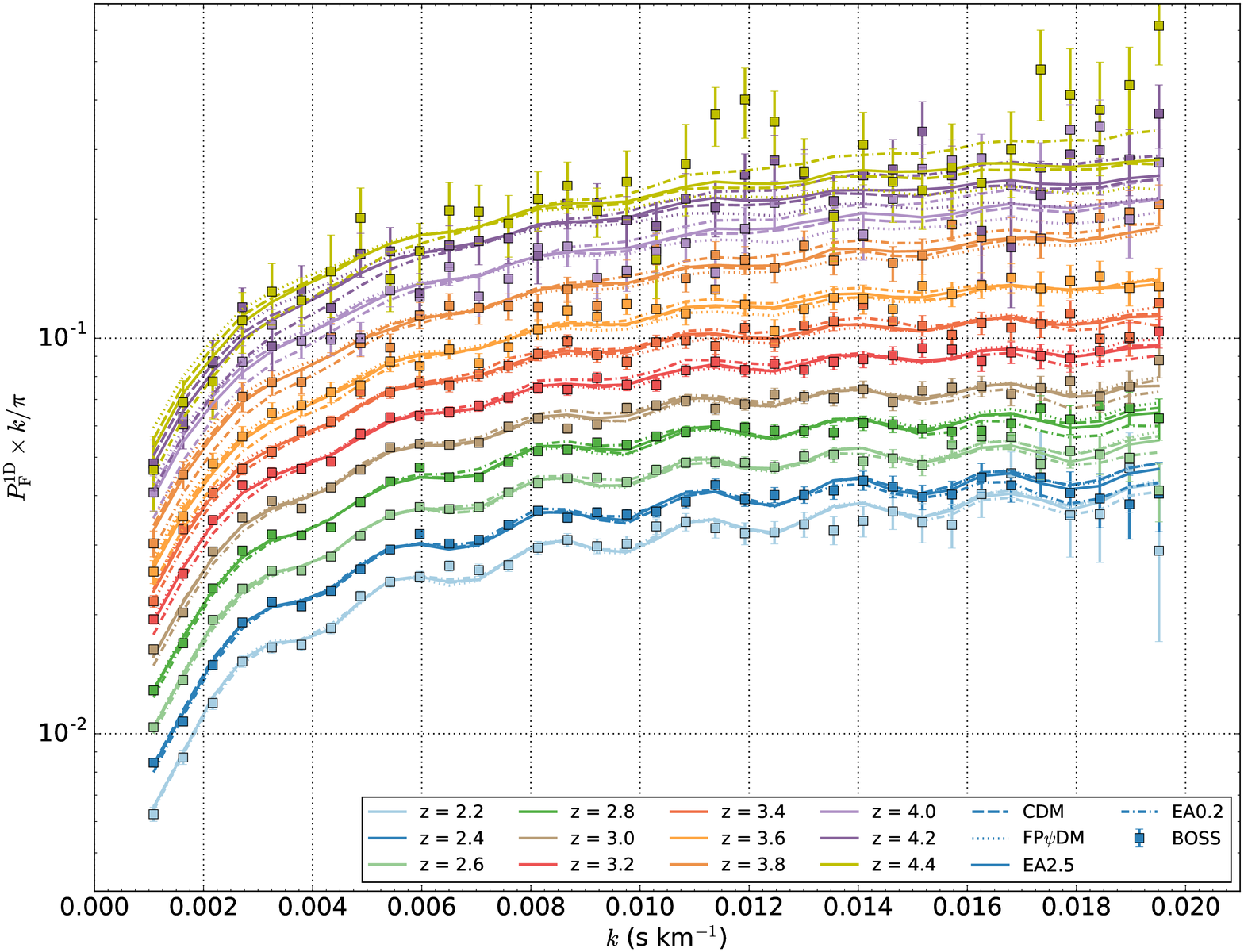}
    \caption{The best-fitted transmitted flux power spectra of CDM, FP$\psi$DM, EA0.2 and EA2.5 and the BOSS data at each redshift bin. All predicted flux power spectra are almost identical for $z<2.6$. The predicted flux power spectra begin to diverge when $z> 3$. In particular the difference among all predicted flux power spectra is significant for $z>3.8$. This suggests the structure evolution is diverging at high redshift but become similar at the low redshift, which is supported by the evolution of matter power spectra of different dark matter models (Fig. \ref{fig:DM_PS_I})}.
    \label{fig:FPS_CDM_EA1_BOSS}
\end{figure*}

\begin{figure}
	\includegraphics[width=\columnwidth]{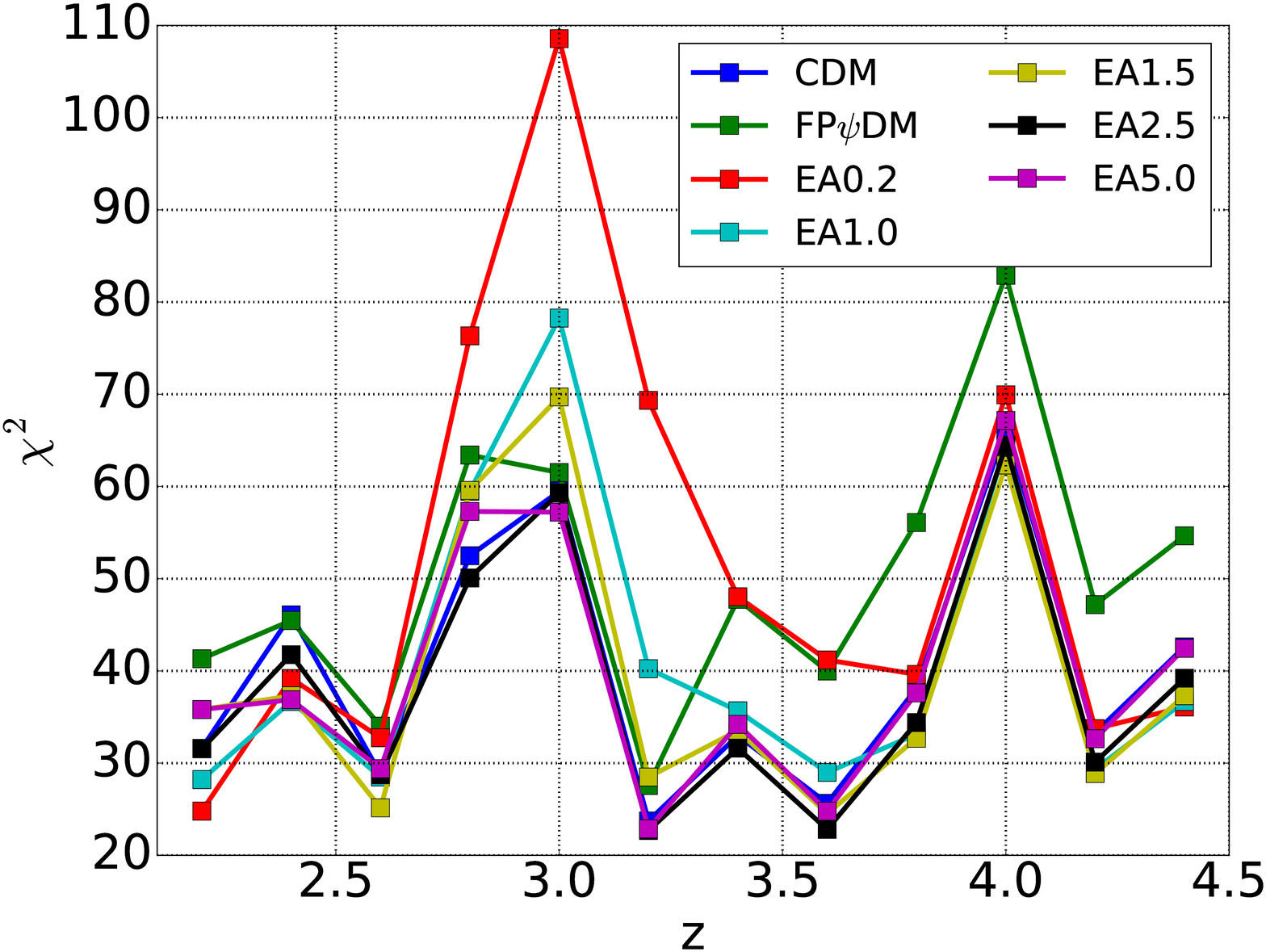}
    \caption{The $\chi^{2}$ distributions of different simulations at each redshift bin.  Two abnormal peaks occur at redshift bins $z=3$ and $z=4$ for all models while other redshift bins appear normally distributed.}
    \label{fig:distribution}
\end{figure}

\label{sec:1DFPS} 
The transmitted flux power spectra provided by \cite{2013A&A...559A..85P} include $35$ k-modes ranging from $k = 0.001$ to 0.02 s km$^{-1}$  in $12$ equally spaced redshift bin from $z = 2.2$ to $4.4$. \cite{2013A&A...559A..85P} selected $13821$ spectra from the DR9 quasar catalogue of BOSS \citep{2012A&A...548A..66P} that have high signal-to-noise ($S/N>2$) and no broad absorption damped Ly$\alpha$ features with detectable Lyman limit systems.  The Ly$\alpha$ forest is defined by the rest-frame interval $1050 < \rm RF < 1180$ \AA. \cite{2013A&A...559A..85P} introduced two methods,  the Fourier transform method and the likelihood method, to compute the flux power spectrum.  Both methods yield compatible results and so we compare our results only with the flux power spectra obtained from the Fourier transform method.  We obtain the best-fit predicted spectrum to the BOSS spectrum by minimising the chi-square. The total chi-square, $\chi^{2}_{\rm total}$, is computed as
\begin{equation}
    \chi^{2}_{\rm total} =\sum_{j=1}^{12}\chi^2(z_j) + \left(\frac{\alpha_{\rm reso}}{5}\right)^{2},
	\label{eq:chi}
\end{equation}
with
\begin{equation}
    \chi^{2}(z_j)\equiv \Delta^{T}_jC_{cov, j}^{-1}\Delta_j +\left(\frac{\alpha_{\rm noise}(z_j)}{0.1}\right)^{2},
	\label{eq:chi_z}
\end{equation}
where  $\chi^2(z_j)$ is the chi-square at $z_{j}$, $C_{cov,j}$ is the covariance matrix of the BOSS data at $z_j$ and the vector $\Delta_j$ is defined as $\Delta(k_i, z_j) = P_{\rm F}^{\rm 1D, fit} (k_{i}, z_{j}) - P_{\rm F}^{\rm 1D, BOSS} (k_{i}, z_{j})$.  We adjust all 26 fitting parameters at the same time for each model fitting to minimize $\chi^{2}_{\rm total}$ through the gradient descent method. 
Fig. \ref{fig:FPS_CDM_EA1_BOSS} shows our best-fit $P_{\rm F}^{\rm 1D, fit} (k)$ of CDM, FP$\psi$DM and EA$\psi$DM with various $\delta\theta$ for all redshifts. As can be seen, all best-fit 1D flux power spectra are comparable and consistent with the BOSS data in the redshift range $z<2.6$. In contrast, all best-fit flux spectra diverge in the range $z = 3.8- 4.4$, indicating that different DM models predict significantly different Ly$\alpha$ forest spectra at high redshifts. A similar phenomenon has been observed for warm dark matter models \citep[][etc]{2013MNRAS.429.1734V,2017PhRvD..96b3522I}. Fig. \ref{fig:distribution} shows the distribution of $\chi^{2}(z_j)$ as functions of redshift for representative models (CDM, FP$\psi$DM, EA0.2 and EA2.5). We note that two $\chi^2(z_j)$ peaks at $z=3$ and $z=4$ for almost all models, and will come back to this issue later.

\begin{figure}
	\includegraphics[width=\columnwidth]{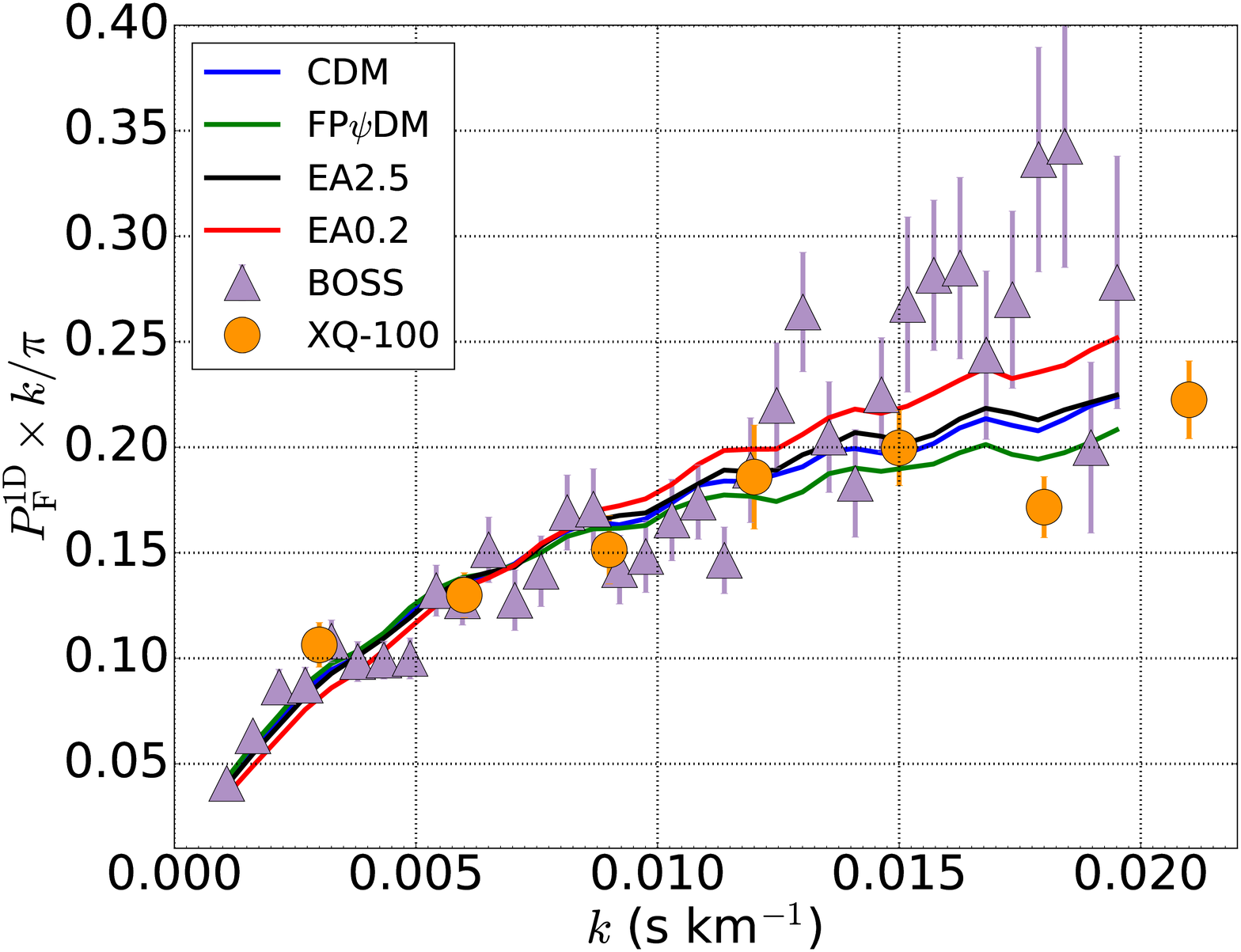}
    \caption{The transmitted flux power spectra of BOSS, XQ-100 and model predictions at $z=4.0$. It clearly shows that the power of BOSS is higher than that of XQ-100 around $k > 0.015$ s km$^{-1}$, while the predictions are closer to the latter than the former.}
    \label{fig:XQ-100}
\end{figure}
\begin{table}
\centering
\caption{Best fit $\chi^2_{total}$'s of all models}
\label{tab:total_chi}
\begin{tabular}{lccr} 
	\hline
	Name of model & $\chi^2_{total}$\\
	\hline
	CDM & 481.1\\	
	EA$\psi$DM ($\delta\theta = 0.2^{\circ}$) & 619.6 \\
    EA$\psi$DM ($\delta\theta = 1.0^{\circ}$) & 499.6\\
    EA$\psi$DM ($\delta\theta = 1.5^{\circ}$) & 475.3 \\
    EA$\psi$DM ($\delta\theta = 2.5^{\circ}$) & 456.8 \\
    EA$\psi$DM ($\delta\theta = 5.0^{\circ}$) & 478.3 \\
    FP$\psi$DM & 601.9 \\
	\hline
\end{tabular}
\end{table}

To estimate the difference between CDM and $\psi$DM, we evaluate the confidence interval (CI), following the frequentist interpretation used in \cite{2015JCAP...02..045P}; \cite{2017MNRAS.471.4606A}. The $\chi^{2}_{\rm total}$ of each dark matter model is shown in Table \ref{tab:total_chi}.
Though the reduced $\chi^2_{\rm total}(=\chi^2_{\rm total}/420$ where $420$ is the total number of data points) of the CDM, EA1.5 and EA2.5 models are comparable and close to unity, the difference between the EA2.5 model and the CDM model is statistically significant. The FP$\psi$DM and CDM models can be considered as the two limits $\delta\theta \to \pi$ and $m_b\to \infty$, respectively. We thus have $\delta\theta$ and $m_b$ as free parameters for all models under test. If the cosmological parameters fixed to the ones optimised for the CDM model throughout all simulations are also the best parameters for EA2.5, which is clearly less optimal, we will have only two degrees of freedom, $N_{\rm dof}= 2$.  Assuming the EA2.5 model is the best model, we can then evaluate the confidence interval, CI, through the following equation:

\begin{equation}
   CI(m_{b}, \delta\theta)=\int_{0}^{\Delta\chi^{2}(m_{b}, \delta\theta)}f_{\chi^{2}}(x; N_{\rm dof})dx,
	\label{eq:CL}
\end{equation}
with
\begin{equation}
   f_{\chi^{2}}(x; N_{dof})=\frac{x^{N_{\rm dof}/2-1}e^{-x/2}}{2^{N_{\rm dof}/2}\Gamma(N_{\rm dof}/2)},
	\label{eq:chi_PDF}
\end{equation}
where $\Gamma$ is the Gamma function. Using these equations and the $\chi^{2}_{\rm total}$ difference between EA2.5 and CDM, $\Delta\chi^{2}_{\rm total} = 24.3$, the CDM model is at the confidence interval, $CI(m_b\to \infty, \delta\theta \to \pi)\simeq 1-5.3\times 10^{-6}$. Hence, the difference between CDM and EA2.5 is exceedingly significant.

Actually, in this work we have not thoroughly optimised the assumed best model, EA2.5, having a fixed $m_b$ (to $1.1 \times 10^{-22}$eV) and only coarsely sampling $\delta\theta$. In addition, the cosmological parameters should have also been optimised for all EA$\psi$DM simulations, but we instead fix these parameters to the one value optimised for the CDM model in favour of CDM. These two factors will affect the above evaluation of CI, but the result is unlikely to change much; moreover, these factors will also change the best values of the two degrees of freedom.  If we set $m_b\sim 10^{-22}$ eV as a prior given by the best fit of $\psi$DM to the Fornax dwarf galaxy data \citep{2014NatPh..10..496S}, the best EA$\psi$DM model can be estimated by interpolation of Table (\ref{tab:total_chi}), and the best $\delta\theta$ lies between $2^o$ and $4^o$.  

\section{Discussions}
\label{sec:discussion}
Fig. \ref{fig:distribution} reveals two distinct peaks in the $\chi^2$ distributions for all dark matter models at the same redshifts. This unusual behaviour suggests that the predicted 1D flux power spectra have non-negligible collective discrepancies with the BOSS data at $z=3.0$ and $z=4.0$ and it calls for further close examinations.

First, the fact that $\chi^2(z)$ gets abruptly enhanced at $z=4$ may have been related to the problem of the BOSS data quality. Fig. \ref{fig:XQ-100} compares the flux power spectra data of BOSS and XQ-100 at $z=4.0$, clearly revealing that the BOSS spectrum is inconsistent with XQ-100 spectrum at high-$k$ bins. Our best-fit spectra tend to follow the XQ-100 spectrum more closely than the BOSS spectrum at high-$k$. We have also checked the consistency between the BOSS data and the XQ-100 data in other redshift bins and found that, except for $z=4.0$, they are all consistent. Therefore, this peak of $\chi^2(z)$ at $z=4$ is likely caused by some systematics of high-$k$ BOSS data at this redshift. 

Second, the absence of non-uniform ionisation of helium gas in our simulations may explain the poor match of our predictions with the data at $z=3$.  Recent observations suggest that the helium reionisation epoch had been started at $z \gtrsim 3.0$ and ended at $z \simeq 2.7$:  (a) a helium Gunn-Peterson trough at $z \simeq 3.0$ was reported by \cite{2014ApJ...784...42S} indicating that reionisation of helium is not completed by that time, but (b) HeIII Ly$\alpha$ has already become transparent at $z \simeq 2.7$ \citep{2011ApJ...733L..24W}. It has been asserted that despite the presence of local UV sources, uniform UVB can be a good approximation for $z\sim 2-4$, except $z=2.7-3$, due to large UV mean free paths.  But during $z \sim 2.7-3.0$, intense UV radiation from the onset of quasars renders the helium rapidly reionised. Helium reionisation reduces the UV mean free path and is patchy, thus yielding local UV heating \citep{2017arXiv171003286L}. Our simulations assume uniform UVB and do not take the non-uniform UV heating into account. This problem possibly leads to the more substantial deviation of our predictions from the BOSS data around $z =3$ than expected.  

\section{Conclusion}
\label{sec:conclusion}
In this paper, we investigate the viability of EA$\psi$DM to explain Ly$\alpha$ forest absorption spectra. Our N-body hydrodynamical simulations in the wave-like dark matter scenario are based on three hypotheses. I) Quantum effects are approximately represented by modification on the linear matter power spectrum. II) UV background is spatially uniform and gas is optically thin and in ionisation equilibrium.  III) The cosmological parameters are the same as the one optimised for the CDM model. Our simulations produce predicted Ly$\alpha$ absorption spectra from different dark matter scenarios upon applying a posterior process discussed in Sec. \ref{sec:Fitting}. Confronting the low-resolution BOSS data, we have approximately identified that the EA$\psi$DM model with $\delta\theta \sim 2.5$ best matches the observation assuming $m_b = 1.1 \times 10^{-22}$ eV. The more precise value of $\delta\theta$ likely lies in the range of $2.5^o - 3.5^o$ through interpolation from Table 1.    

Our results further show that the predicted Ly$\alpha$ flux power spectra in the CDM model produces a significantly larger $\chi^2_{\rm total}(=481)$ than the best EA$\psi$DM model of $(m_b, \delta\theta) = (1.1 \times 10^{-22}$eV$, 2.5^o)$ does with $\chi^2_{\rm total}=456.8$. Though all cosmological parameters used in EA$\psi$DM simulations are optimised for CDM, the best EA$\psi$DM model still provides the smallest $\chi^2_{\rm total}$. The difference between the best EA$\psi$DM model and the CDM model is statistically significant and the CDM model is outside the confidence interval $CI \sim 1-10^{-5}$, assuming EA2.5 is the best model. 

High-resolution data, such as XQ-100 and HIRES/MIKE, can provide stronger constraints on the axion mass and the axion angle. (See the demonstration in Appendix \ref{sec:spectra}.) However, as discussed in Sec. \ref{sec:Matter_power_spectrum}, quantum effects become important when the spectral resolution is as high as these data. These effects are beyond the capability of the N-body hydro simulation presently employed and can only be reliably captured in the wave simulations, which await future investigations.  In addition, the stellar feedback discussed below may no longer be neglected in the simulation to compare with the high-resolution data, and accurate small-scale hydrodynamics modeling is thus needed.

Most Ly$\alpha$ simulation works do not take into account mechanical feedback from stars, which is clearly an important source for IGM turbulence and heating.  This issue may cause concerns for the credibility of the predicted flux spectrum.  Here we present an argument in favour of the BOSS data.  The BOSS data have the highest resolution about $k_{\rm Ny}\sim 5$ $h$Mpcc$^{-1}$ or a physical wavelength $200$ kpc around $z\sim3$.  Stellar mechanical feedback is unlikely to reach this large scale and only AGN feedback is possible to create an impact on high-$k$ spectra of BOSS. But AGNs are rare and moreover their activities peak at $z=2$ which is after the lowest redshift of the BOSS data at $z=2.2$.  Hence the BOSS data are free from the contamination of the stellar feedback and marginally free from the AGN feedback. 

We have assumed in this work uniform UV backgrounds which give rise to global heating.  But local UV heating can also be significant, especially around $z=2.7-3$, and such local UV heating has not been included in our simulations. Unlike local stellar feedbacks, local UV sources from quasars,
which reionise helium thereby heating the gas near $z=3$,
can have an impact on scales spanned by the BOSS data.  Despite UV sources are non-trivial to model, some simple heating recipe can indeed bring the predicted flux spectrum in closer agreement with the BOSS data.  As demonstrated in \citep{2017MNRAS.471.4606A}, incorporating empirical local UV heating yields $\chi^2_{\rm total,CDM}=405$ rather than $\chi^2_{\rm total,CDM}=481$ in our prediction. Although some fitting parameters in that work are different from ours which may affect $\chi^2_{\rm total}$, the $\chi^2$ peak around $z=3$ in Fig. \ref{fig:distribution} has a major contribution to the increase of $\chi^2_{\rm total,CDM}$ in our prediction.  This aspect needs improvements in the future work. We also notice that both the CDM model and the EA2.5 model have almost identical large $\chi^2$'s at $z=2.8$ and $z=3$ bins. Affected only by the gas physics rather than underlying dark matter models, in a work with proper local heating to bring down $\chi^2$'s of these two bins, the revised values of $\chi^2$'s in both models are likely comparable. Hence, the difference of $\chi^2_{\rm total}$'s in these two models would not be much affected.

Recent studies on the predicted Ly$\alpha$ flux spectra of free-particle (FP) $\psi$DM placed a lower bound for the particle mass, $m_b \gtrsim 2.9\times10^{-21}$ eV \citep{2017MNRAS.471.4606A} and $m_b \gtrsim 3.75 \times 10^{-21}$ eV \citep{2017PhRvL.119c1302I}\citep[see also][]{2018MNRAS.478.3935N}, which are in tension with the particle mass determined from local satellite galaxies \citep{2014NatPh..10..496S,2017MNRAS.468.1338C}, $m_b\sim 1-3 \times 10^{-22}$ eV.  This work demonstrates that such tension can be removed when the extreme axion misaligned angle is taken into account, i.e., EA$\psi$DM models, even for the lowest particle mass $m_b\sim 10^{-22}$ eV determined from local galaxy data. Due to the lack of local UV heating and various physical effects not taken in account in this demonstrative work, we do not obtain a $\chi^2_{\rm total}$'s as small as they should have been and cannot ascertain the exact values of optimal axion angle $\delta\theta$ and particle mass $m_b$.  But the tendency of a possible superior extreme axion wave dark matter model over, or at worst comparable to, the CDM model against the BOSS Ly$\alpha$ forest data is evident.  

Finally, we would like to put the finding of the best initial axion angle $\theta_0$ into a perspective of the axion model.  In the axion model, $f$, the decay constant, also characterises the axion symmetry-breaking energy scale, and $f$ is on one hand determined by matching the matter energy density $m_b^2 c^2 f^2 \theta^2$ equal to the radiation energy density at the radiation-matter equality.  Well prior to the equality, $\theta^2$ is decreasing with the scaling factor as $(a/a_{eq})^{-3}$, where $a_{eq}$ is the scaling factor at the equality, and the matter energy density behaves as CDM does.  But this $a^{-3}$ scaling occurs only after the Zitterbewegun oscillation starts at $a_{osc}$; before the oscillation the matter energy density is a constant.  As a result, the matter energy density at the very early epoch on the other hand is $m_b^2c^2 f^2\theta_0^2 \sim \epsilon_{rad, eq} (a_{eq}/a_{osc})^3$, where $\epsilon_{rad, eq}$ is the radiation energy density at the equality.  The squared decay constant $f^2=\epsilon_{rad, eq}(a_{eq}/a_{osc}(\theta_0))^3/m_b^2 c^2\theta_0^2$, and this quantity turns out to be a decreasing function of $\theta_0$.  For example, the FP$\psi$DM model, where $\theta_0\to 0$, has the axion symmetry-breaking scale $f \to \theta_0^{-1}$, and the symmetry-breaking scale may exceed the Planck scale.  For $m_b=1.1\times 10^{-22}$eV and $\delta\theta_0 = 5^{\circ}, 2.5^{\circ}, 1.5^{\circ}$, the three initial angles found to have smaller $\chi^2_{total}$ than the CDM model, we have the symmetry-breaking scale $f = 2.66, 2.32$, and $2.13 \times 10^{16}$ GeV, slightly above the typical grand unified theory energy scale $10^{16}$ GeV, which is also the fiducial energy scale of the inflation.  The axion symmetry-breaking energy scale above or equal to the inflation energy scale has an important merit that topological defects, such as domain walls arising from different parts of the universe having different values of $\theta_0$'s can be inflated away, therefore ensuring that we have one and only one value of $\theta_0$ in the visible universe, as assumed in this work.  Interestingly, if we let $m_b$ to vary and $m_b > 10^{-22}$ eV, all axion symmetry-breaking scale should exceeds the fiducial inflation scale.  This is because $a_{osc}\propto m_b^{-1/2}g(\theta_0)$ for some function $g$ of $\theta_0$, we find $f$ only weakly depends on $m_b$ as $\propto m_b^{-1/4}g^{3/2}(\theta_0)\theta_0^{-1}$, and both $g(\theta_0)$ and $\theta_0^{-1}$ are increasing functions of $m_b$ when the optimal $\theta_0$ for the Ly$\alpha$ data is taken into account, which compensate the weak dependence of $m_b^{-1/4}$.



\section*{Acknowledgements}
We thanks Dr. Phil Hopkins making the GIZMO code available to us. T.C. acknowledges the funding support from MOST of Taiwan under the grant, MOST 107-2119-M-002-036-MY3.




\bibliographystyle{mnras}
\bibliography{example} 




\appendix

\section{Splicing method}
\label{sec:Splicing}
In Sec. \ref{sec:Mockspectra}, we mentioned that we simulated $P^{\rm 1D}_{\rm F}$ through the splicing method \citep{2003ApJ...585...34M,2014JCAP...07..005B} to reduce the computation time for hydrodynamical simulations. In practice, different simulations with varying resolutions and with varying box sizes provide different dynamical ranges of $k$.  In this appendix, we present our tests of the splicing method.

The lowest $k$ and the highest $k$ of a simulation with $(L, N_{\rm gas})$ are $k_{\rm min,L} = 2\pi/L$ and $k_{\rm max} = k_{\rm Ny,L} = \pi N_{\rm gas}^{1/3}/L$, respectively, where $k_{\rm Ny,L} $ is the Nyquist wavenumber. We test the splicing method by considering three $k$ regimes, each corresponding to the range, $k<k_{\rm min, 25}$,  $k_{\rm min, 25}<k<k_{\rm Ny, 100}/4$ and $k>k_{\rm Ny, 100}/4$, respectively, where $k_{\rm min, 25} (=  2\pi/25$ $h^{-1}$Mpcc) is the minimum k for $L= 25 $ $h^{-1}$Mpcc and $k_{\rm Ny, 100}( = 512\pi/100$ $h^{-1}$Mpcc) the Nyquist wavenumber of $(100$ $h^{-1}$Mpcc , $2\times512^{3})$ simulation. For the range $k<k_{\rm min, 25}$, $P_{\rm F,100,512}^{\rm 1D} (k)$ is considered as a reference and $P_{\rm F,25,512}^{\rm 1D} (k_{\rm min, 25})/P_{\rm F,25,128}^{\rm 1D} (k_{\rm min, 25})$ as k-independent correction factor, and the splicing method is expressed as:
\begin{equation}
    P_{\rm F}^{\rm 1D} (k)= P_{\rm F,100,512}^{\rm 1D} (k)\times\frac{P_{\rm F,25,512}^{\rm 1D} (k_{\rm min, 25})}{P_{\rm F,25,128}^{\rm 1D} (k_{\rm min, 25})}.
    	\label{eq:Splicing_II}
\end{equation}
For $k_{\rm min, 25}<k<k_{\rm Ny, 100}/4$, the effective $k$ ranges of each simulation are overlapped and the splicing method is expressed as:
\begin{equation}
    P_{\rm F}^{\rm 1D} (k)= P_{\rm F,100,512}^{\rm 1D} (k)\times\frac{P_{\rm F,25,512}^{\rm 1D} (k)}{P_{\rm F,25,128}^{\rm 1D} (k)}.
	\label{eq:Splicing_III}
\end{equation}
For $k>k_{\rm Ny, 100}/4$, $P_{\rm F,25,512}^{1D} (k)$ is considered as a reference and $P_{\rm F,100,512}^{\rm 1D} (k_{\rm Ny, 100}/4)/P_{\rm F,25,128}^{\rm 1D} (k_{\rm Ny, 100}/4)$ as k-independent correction factor, and the splicing method is expressed as:
\begin{equation}
    P_{F}^{1D} (k)= P_{F,25,512}^{1D} (k)\times\frac{P_{F,100,512}^{1D} (k_{Ny, 100}/4)}{P_{F,25,128}^{1D} (k_{Ny, 100}/4)}.
	\label{eq:Splicing_IIII}
\end{equation}

\begin{figure}
	\includegraphics[width=\columnwidth]{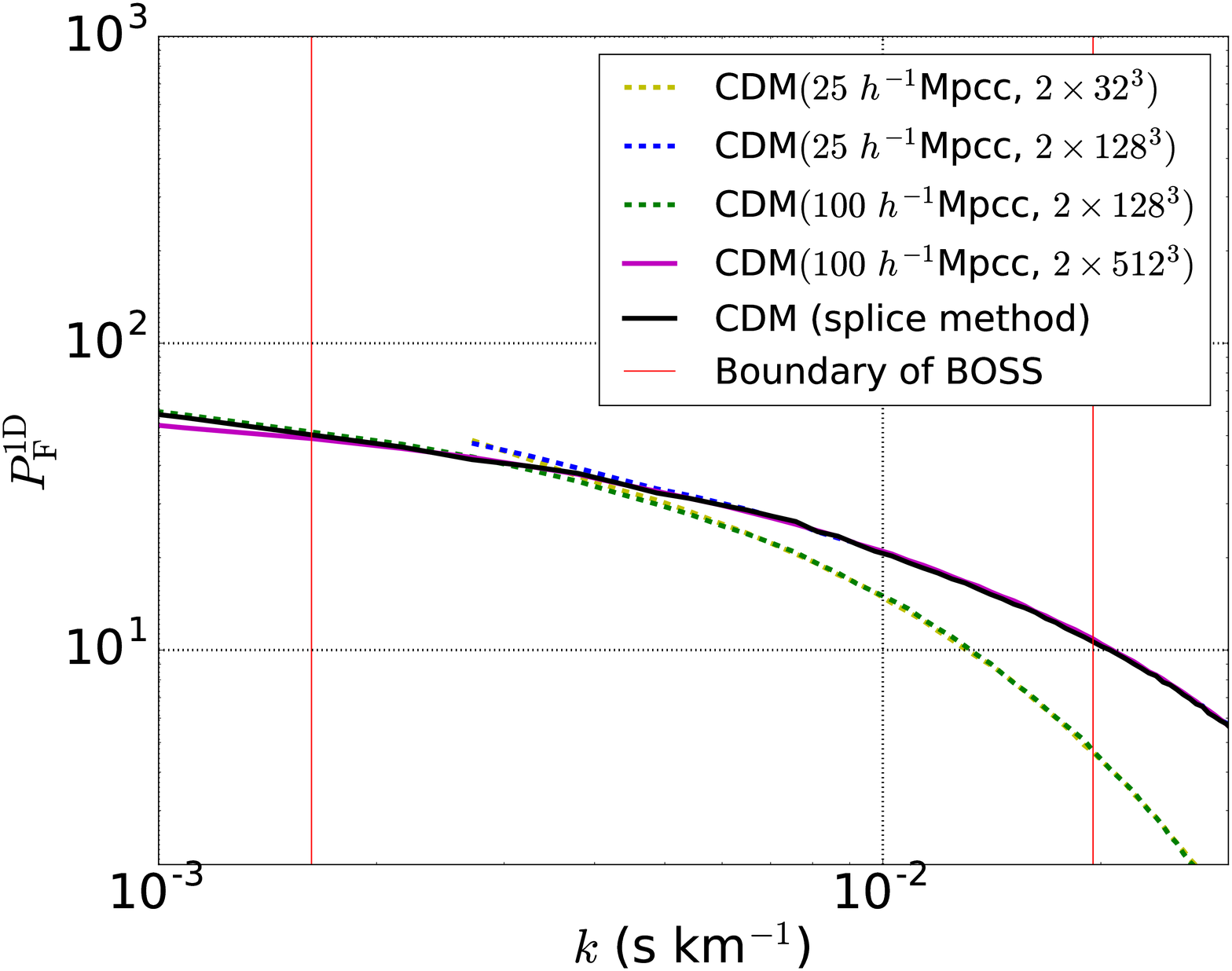}
    \caption{Test of the splicing method for the CDM model. The pink line is the exact transmitted flux power spectra from $(100$ h$^{-1}$Mpcc, $2\times512^{3})$ simulation and the red line is the spliced transmitted flux power spectra composite flux power spectra from $(100$ h$^{-1}$Mpcc, $2\times128^{3})$, $(25$ h$^{-1}$Mpcc, $2\times128^{3})$ and $(25$ h$^{-1}$Mpcc, $2\times32^{3})$ simulations. The vertical lines are the range of the BOSS data in use.}
    \label{fig:Splicing_method}
\end{figure}

We then construct the composite flux power spectrum with $(L, N) =  (100$ $h^{-1}$Mpcc , $2\times512^{3})$ obtained by merging the three  flux spectra of $(L, N) =  (25$ $h^{-1}$Mpcc , $2\times32^{3})$, $(L, N) =  (25$ $h^{-1}$Mpcc , $2\times128^{3})$ and $(L, N) =  (100$ $h^{-1}$Mpcc , $2\times128^{3})$, and compare the composite flux spectrum with the exact flux power spectra of $(L, N) =  (100$ $h^{-1}$Mpcc , $2\times512^{3})$. In Fig. \ref{fig:Splicing_method}, we illustrate the goodness of splicing at $z=3.0$ for the CDM model. The error of splicing method is at most $8\%$ at every redshift over the entire $k$-range of interest.  The errors in other redshifts are comparable to this characteristic value.

\section{Numerical Convergence}
\label{sec:convergence}
\begin{figure}
	\includegraphics[width=\columnwidth]{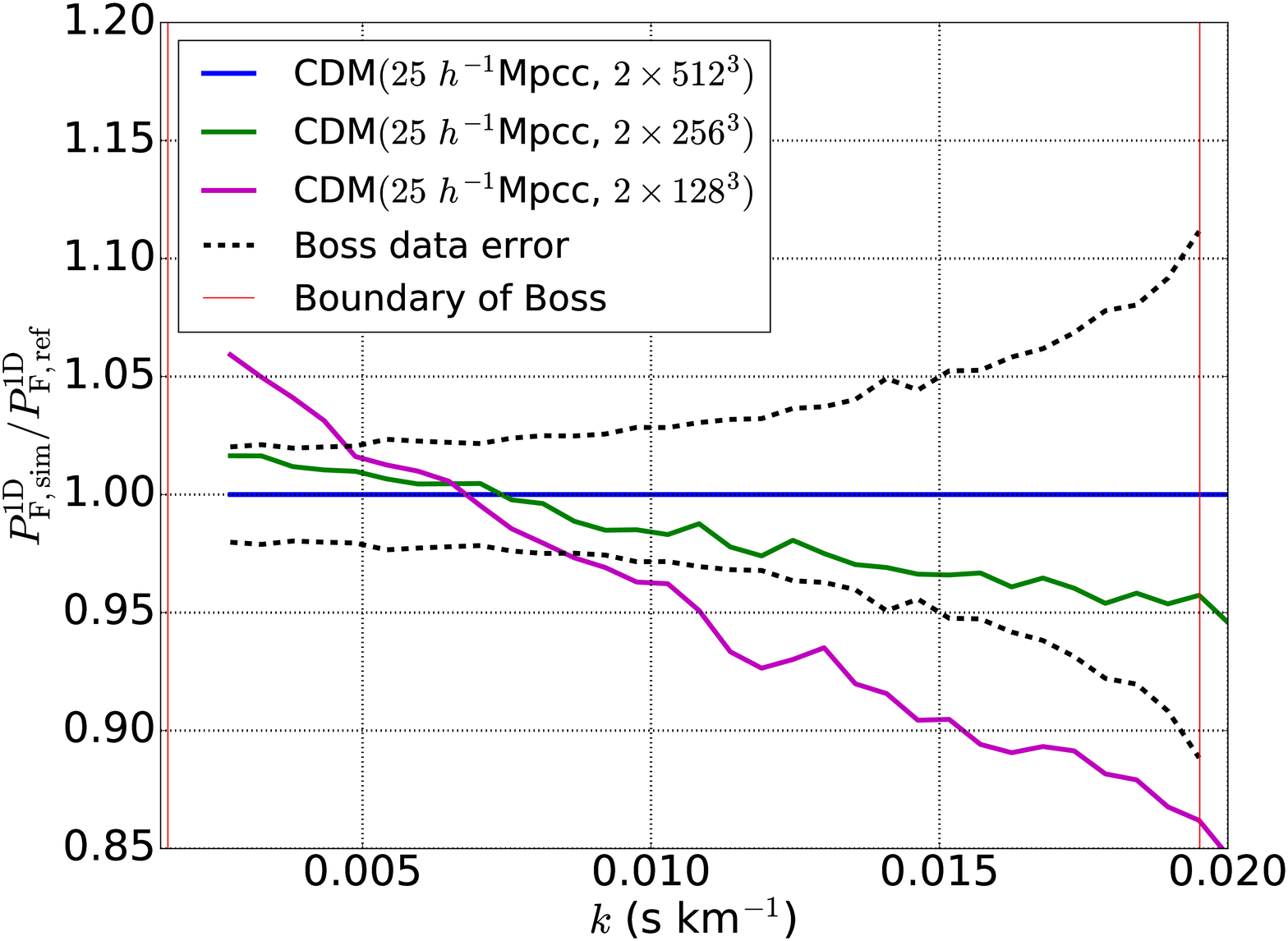}
    \caption{Convergence test in the CDM at $z = 3.0$, where we choose the transmitted flux power spectrum of the $(25$ h$^{-1}$Mpcc, $2\times512^{3})$ simulation as the reference. The blue line,  the green line and the pink line are the transmitted flux power spectra from $(25$ h$^{-1}$Mpcc, $2\times512^{3})$,  $(25$ h$^{-1}$Mpcc, $2\times256^{3})$ and $(25$ h$^{-1}$Mpcc, $2\times128^{3})$ simulations, respectively.  The black dash line represents the normalized $1\sigma$ uncertainties measured in the BOSS data and the vertical red lines are the range of the BOSS data in use.}
    \label{fig:Flux_convergence}
\end{figure}

As our $\psi$DM model simulations are based on the CDM-hydro simulation with modified initial matter power spectra (see Sec. \ref{sec:Hydrodynamical}), the convergence test for the CDM simulation alone should adequately reflect the convergence behaviours of the simulated flux power spectra of all dark matter models.  We choose $P_{\rm F,25, 512}^{\rm 1D}(k)$ as the reference spectrum and define $(\pm \sigma_{\rm BOSS}(k) / P_{\rm F,25, 512}^{\rm 1D}(k)) + 1$ as the normalized data uncertainty, where $\sigma_{\rm BOSS}$ is the $1-\sigma$ uncertainties of the BOSS data. 
We show the convergence test at $z = 3.0$ in Fig. \ref{fig:Flux_convergence}, where the relative flux power spectra of $(25$ h$^{-1}$Mpcc, $2\times128^{3})$ and $(25$ h$^{-1}$Mpcc, $2\times256^{3})$ simulations are shown along with the normalized data uncertainty.  It is clear that the $(25$ h$^{-1}$Mpcc, $2\times256^{3})$ flux power spectrum is within $1 \sigma_{\rm BOSS}$ and converges to within $\sim 5$ percent of the reference spectrum.  Linearly extrapolating $P_{\rm F,\rm sim}^{\rm 1D}(k) / P_{\rm F,\rm ref}^{\rm 1D}(k) - 1$ in log-space, it follows that the difference between the flux power spectra of $(25$ h$^{-1}$Mpcc, $2\times512^{3})$ and $(25$ h$^{-1}$Mpcc, $2\times1024^{3})$ simulations is within $0.2 \sigma_{\rm BOSS}$. This amount of simulation errors may somewhat enhance the total $\chi^2$ listed in Table 1, but a quantitative 
estimate of the enhancement is nontrivial to assess.

\section{Sample Ly$\alpha$ Absorption Spectra}
\label{sec:spectra}
\begin{figure}
	\includegraphics[width=\columnwidth]{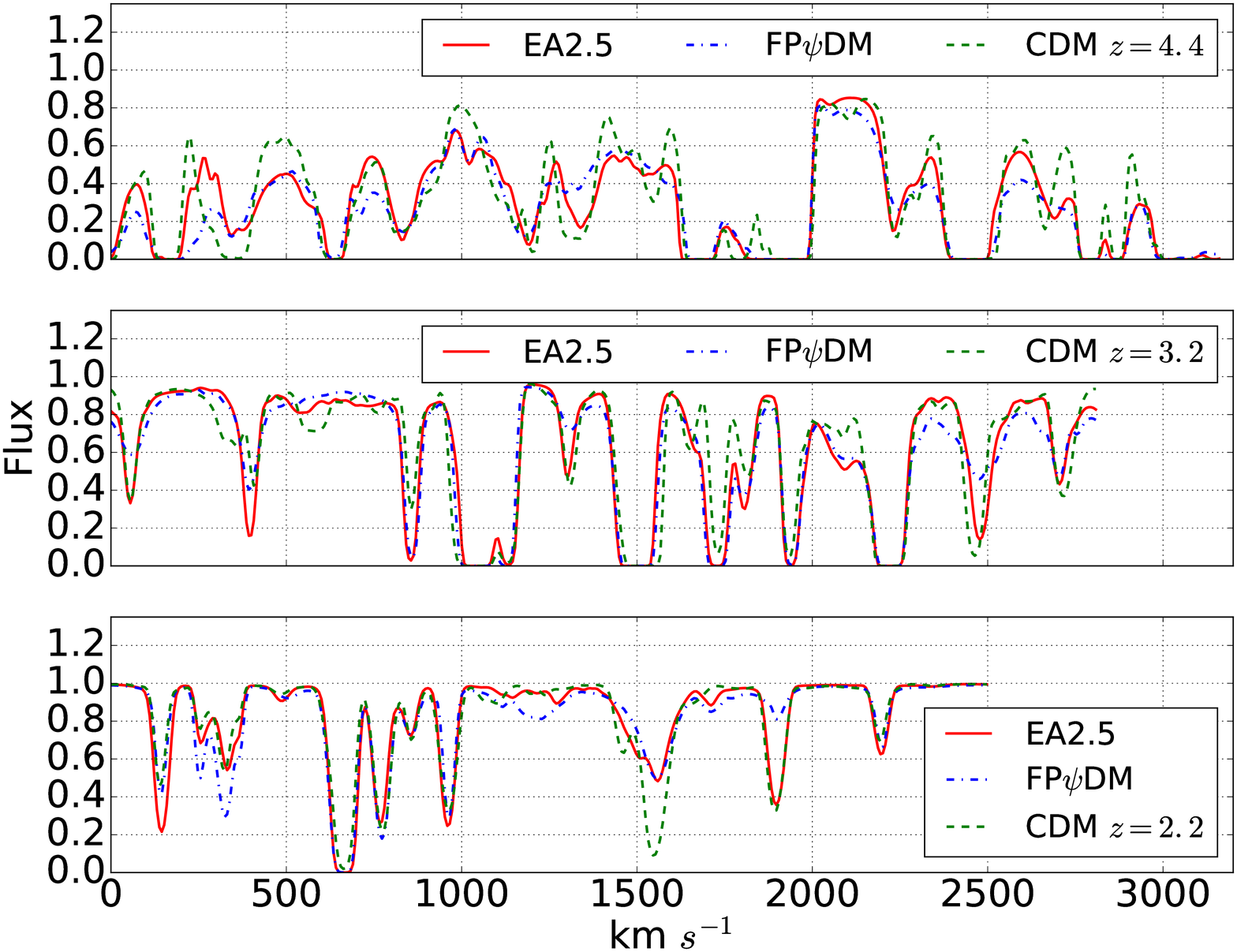}
    \caption{Simulated spectra for EA2.5, FP$\psi$DM and CDM at $z=2.2, 3.2, 4.4$.}
    \label{fig:absorption_spectra}
\end{figure}

We here show a set of sample simulated spectra ($z=2.2, 3.2$ and $4.4$) of different models along the same line of sight in Fig. \ref{fig:absorption_spectra}, with which Fourier transform is performed to obtain flux power spectra and the average power spectrum over $10^5$ samples is to be compared with the BOSS data.  From individual Ly$\alpha$ spectra at $z=2.2$ and $3.2$, it is difficult to detect any systematic difference among different models, partly because the lack of neutral hydrogen at these redshifts
renders the absorption feature less prominent and partly because different models have already been highly evolved which erases the feature of the initial matter power spectrum.  By contrast at
$z=4.4$, the substantially more abundant neutral hydrogen makes the absorption feature more distinct and the dark matter is less evolved than low redshifts.  We can clearly see that the CDM model has large variations in the absorption spectrum than the two $\psi$DM models, where low absorption regions tend to have lower absorption and high absorption regions tend to have higher absorption, indicative of a more clumpy universe for the CDM model.  It also demonstrates that high-$z$ Ly$\alpha$ forest data are crucial for probing the difference in initial power spectra of different models. 

\bsp	
\label{lastpage}
\end{document}